\documentclass{llncs}
\usepackage{makeidx}  
\usepackage{subfig}
\usepackage{graphicx}
\usepackage{amssymb,amsmath,bm}
\usepackage{booktabs}
\usepackage[misc]{ifsym}
\usepackage{multirow}
\usepackage{bbding}
\usepackage{array}
\usepackage{float}
\usepackage{algorithm}
\usepackage{algorithmic}
\usepackage{setspace}

\newcolumntype{C}[1]{>{\centering\let\newline\\\arraybackslash}m{#1}}
\begin{document}
	\frontmatter          
	%
	%
	\mainmatter              
	\title{Multi-sequence Cardiac MR Segmentation with Adversarial Domain Adaptation Network}	
    \author{Jiexiang Wang\thanks{~ indicates equal contributions.} \and Hongyu Huang$^{\star}$ \and Chaoqi Chen \and Wenao Ma \and Yue Huang \and Xinghao Ding\thanks{~ corresponding author.}}
    \institute{School of Information Science and Engineering, Xiamen University, China\\
    	\email{\{wangjx, huanghy, cqchen94, wenaoma\}@stu.xmu.edu.cn}\\
    	\email{\{yhuang2010, dxh\}@xmu.edu.cn}}
	\maketitle
	\date{}
	\hyphenpenalty=5000
	\tolerance=1000
	
\begin{abstract}
Automatic and accurate segmentation of the ventricles and myocardium from multi-sequence cardiac MRI (CMR) is crucial for the diagnosis and treatment management for patients suffering from myocardial infarction (MI). However, due to the existence of domain shift among different modalities of datasets, the performance of deep neural networks drops significantly when the training and testing datasets are distinct. In this paper, we propose an unsupervised domain alignment method to explicitly alleviate the domain shifts among different modalities of CMR sequences, \emph{e.g.,} bSSFP, LGE, and T2-weighted. Our segmentation network is attention U-Net with pyramid pooling module, where multi-level feature space and output space adversarial learning are proposed to transfer discriminative domain knowledge across different datasets. Moreover, we further introduce a group-wise feature recalibration module to enforce the fine-grained semantic-level feature alignment that matching features from different networks but with the same class label.
We evaluate our method on the multi-sequence cardiac MR Segmentation Challenge 2019 datasets, which contain three different modalities of MRI sequences. Extensive experimental results show that the proposed methods can obtain significant segmentation improvements compared with the baseline models.
\end{abstract}
	\section{Introduction}
Accurate segmentation of the ventricles and myocardium is fundamental to the diagnosis and treatment of myocardial infarction (MI) \cite{zhuang2018multivariate}. Cardiac MRI sequences are usually used for the MI diagnosis, in particular the T2-weighted MRI detect damaged and ischemic areas, the balanced-Steady State Free Precession (bSSFP) MRI clearly shows the heart structure boundary, and the late gadolinium enhancement (LGE) MRI can enhance infarcted myocardium with distinctive brightness compared to healthy structure \cite{zhuang2016multivariate}. Manual segmentation is time-consuming, so automatic segmentation is significant in the clinic. Recently, deep learning network has become a powerful tool for
	\begin{figure}[h]
	    \centering
	    \subfloat[\small Source1]{\includegraphics[width=0.15\textwidth]{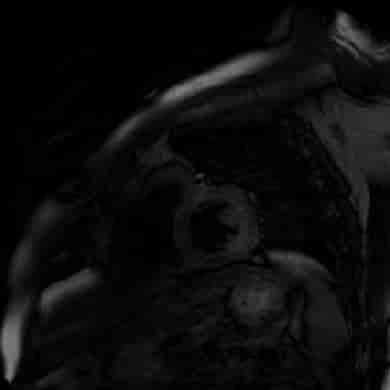}}
	    \hspace{0.3pt}
	    \subfloat[\small Source2]{\includegraphics[width=0.15\textwidth]{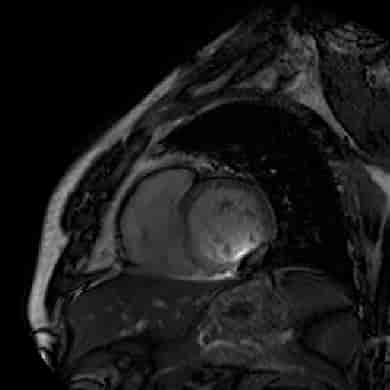}}		
	    \hspace{0.3pt}
	    \subfloat[\small Target]{\includegraphics[width=0.15\textwidth]{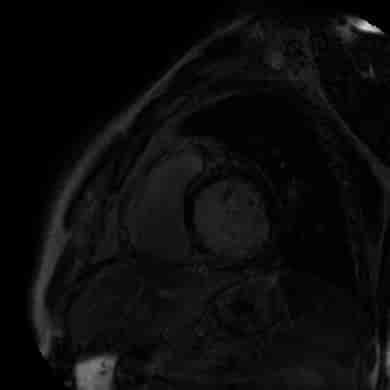}}	
	    \hspace{0.3pt}
	    \subfloat[\small Label]{\includegraphics[width=0.15\textwidth]{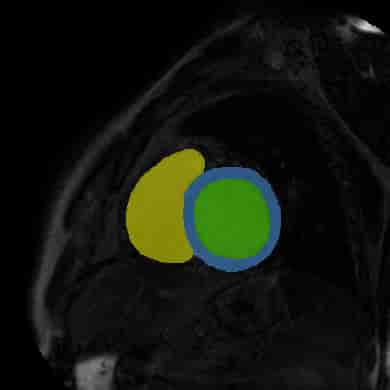}}
	    \hspace{0.3pt}
	    \subfloat[\small T-noDA]{\includegraphics[width=0.15\textwidth]{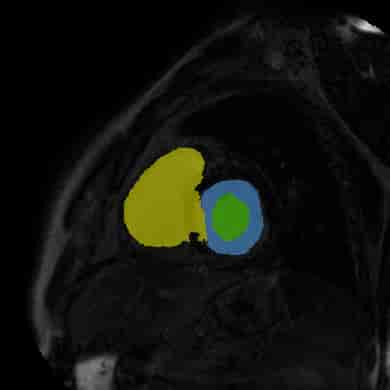}}
	    \hspace{0.3pt}
	    \subfloat[\small T-DA]{\includegraphics[width=0.15\textwidth]{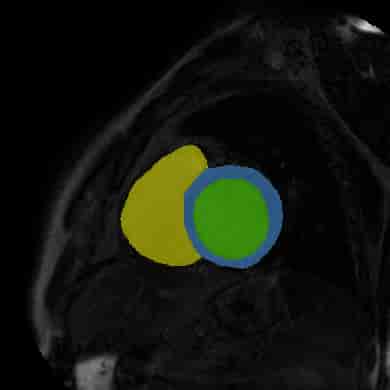}} 	
	    \vspace{-5pt}
	    \caption{\small Performance drops due to domain shift. (a) Original T2-weighted MRI (Source1). (b) Original bSSFP MRI (Source2). (c) Original LGE MRI (Target). (d) LGE MRI annotation (Label). (e) The segmentation results of LGE MRI using an established model trained on T2-weighted and bSSFP MRI data (T-noDA). (f) The segmentation results of an LGE MRI using our model trained on T2-weighted and bSSFP MRI data (T-DA). The yellow region denotes the right ventricle, the green region denotes the left ventricle, and the blue region   denotes the myocardium.}
	    \label{fig:performancedrop}
    \end{figure}semantic segmentation on heart structures \cite{yang2018combating}\cite{yue2019cardiac}. Obviously, the ventricles and myocardium segmentation results can be improved combining the complimentary information from T2-weighted and bSSFP MRI sequences \cite{zhuang2016multivariate}. To save labeling time, sometimes only the T2-weighted and bSSFP MRI sequences and corresponding labels are available. However, a well-trained segmentation model may underperform when being tested on data from different modalities, which is caused by the domain shift (as shown in Fig.~\ref{fig:performancedrop}). Fine-tuning on the target domain data is a simple but efficient method to alleviate the performance drop. But it still requires massive data collection and enormous annotation workload which are impossible for many real-world medical scenarios. For this reason, constructing a general segmentation model suitable for various modalities is promising yet still challenging.

Unsupervised Domain Adaptation (UDA) methods have shown compelling results on reducing the dataset shift across distinct domains. Prior efforts on this problem intended to match the source and target data distributions to learn a domain-invariant representation. For example, Maximum Mean Discrepancy (MMD) was introduced to minimize the distance of source and target feature distributions in Reproducing Kernel Hilbert Space (RKHS)~\cite{tzeng2014deep}. CycleGAN~\cite{zhu2017unpaired} tackled the image-to-image translation task in a fully unsupervised manner, and thus is capable of reducing the domain shift in the pixel-level. AdaptSegNet~\cite{tsai2018learning} solved the unsupervised cross domain segmentation problem by leveraging the domain adversarial training approach. In the context of medical imaging, \cite{dong2018unsupervised} developed an UDA framework based on adversarial networks for lung segmentation on chest X-rays. \cite{ren2018adversarial} improved the UDA framework with Siamese architecture for Gleason grading of histopathology tissue. \cite{dou2018unsupervised} proposed a domain critic module and a domain adaptation module for the unsupervised cross-modality adaptation problem. These approaches, which based on the domain adversarial training, required empirical feature selection. \cite{chen2019synergistic} proposed the synergistic fusion of adaptations from both image and feature perspectives for heart structures segmentation. However, this approach, which based on image-to-image adaptation, cannot be directly introduced to the multiple source domain adaptation problems due to the presence of multiple domain shifts between different source domains.

In this paper, we propose a domain alignment method for the UDA problem, which helps the established model segment the ventricles and myocardium accurately in the target domain without requiring target labels. Firstly, in order to reduce the domain shift with respect to the image appearance, we propose a histogram match operation for all the data. Secondly, we introduce the domain adversarial training in the output space, which can directly align the predicted segmentation results across different domains. Finally, we further propose a group-wise feature recalibration module (GFRM) to improve the domain adversarial training by integrating multi-level features without requiring manual selection to progressively align the source and target feature distributions. The proposed method is extensively evaluated on the multi-sequence cardiac MR Segmentation (MS-CMRSeg) Challenge 2019 datasets, including bSSFP, LGE and T2-weighted MRI sequences.

	\vspace{-0.3cm}
	\section{Method}
\begin{figure}[H]
	    \centering
	    \subfloat{\includegraphics[width=1.0\textwidth]{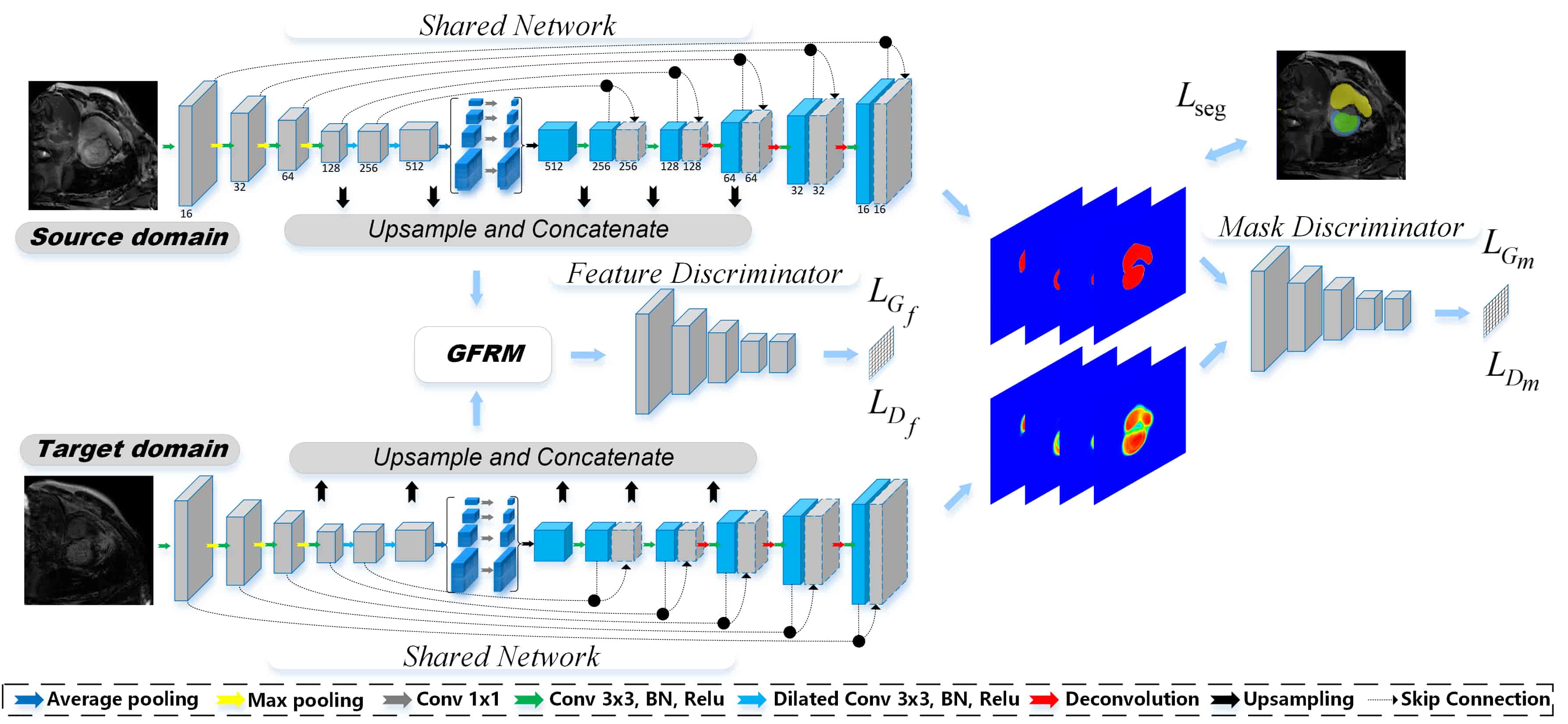}}
	    \caption{Schematic view of our proposed framework.} 
	    \label{fig:wholenetwork}
    \end{figure}

	Fig.~\ref{fig:wholenetwork} overviews our segmentation method for ventricles and myocardium in MRI sequences. We use modified 2D attention U-Net with pyramid pooling module as our segmentation backbone architecture \cite{oktay2018attention}\cite{zhao2017pyramid}. To align the distance over feature and output spaces across different domain, feature-level and mask-level discriminator are adopted. Moreover, the group-wise feature recalibration module (GFRM) is introduced to transfer multi-level feature information. The details of the above modules are shown in Fig.~\ref{fig:subnetwork}. 
	\vspace{-0.5cm}
    \vspace{-0.8cm}	

	\subsection{Network Architecture}
\begin{figure}[H]
		\centerline{\includegraphics[width=0.95\textwidth]{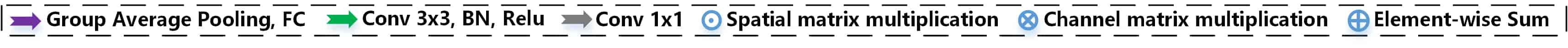}}
    	\hspace{2pt}
        \subfloat[]{\includegraphics[width=0.32\textwidth]{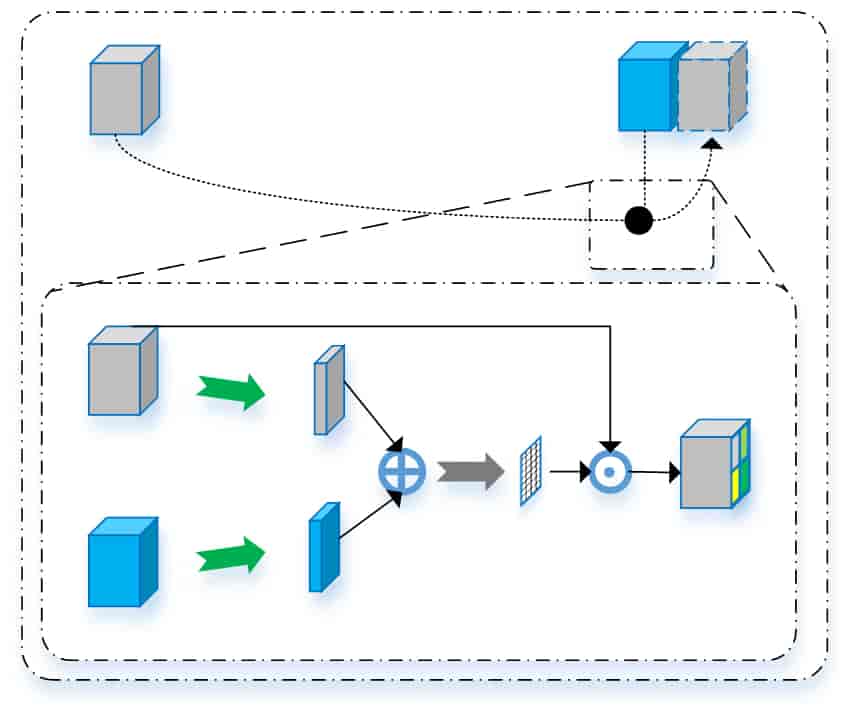}}
        \hspace{1pt}
        \subfloat[]{\includegraphics[width=0.64\textwidth]{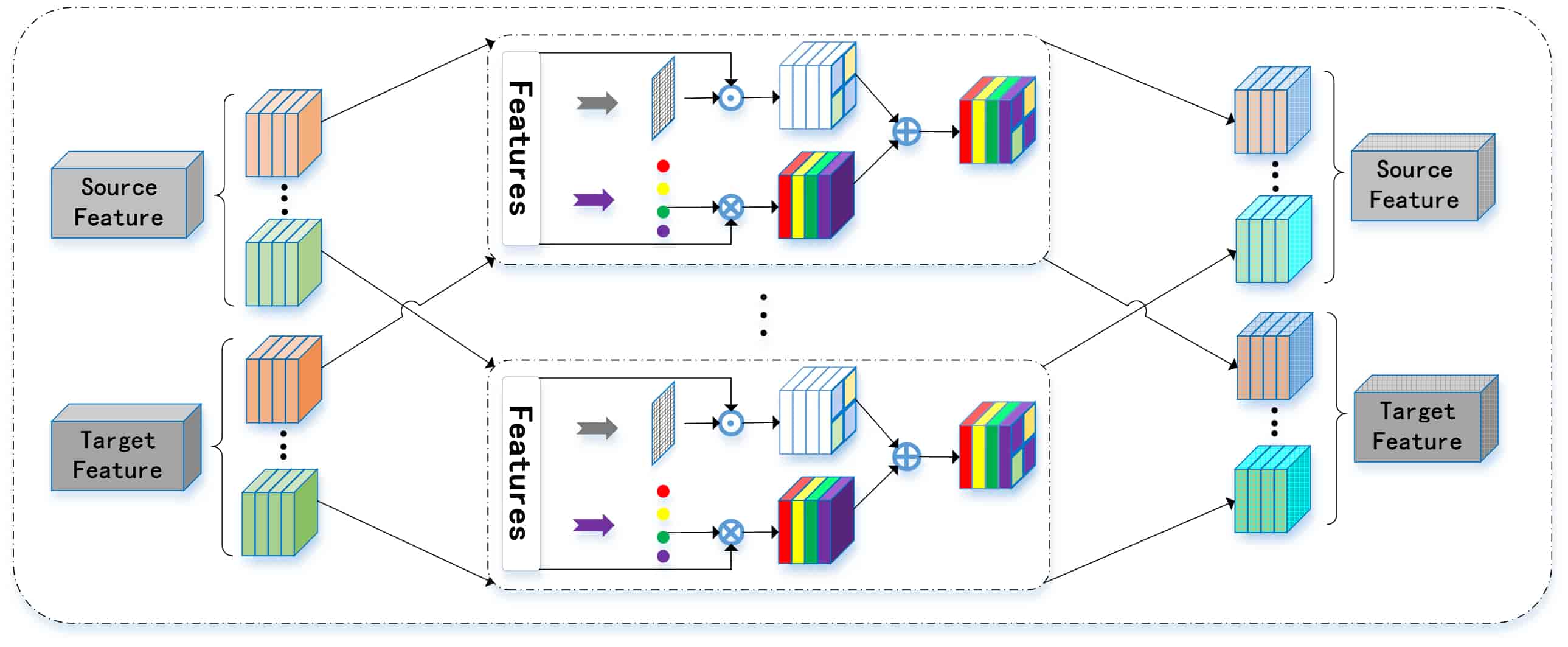}}
        \hspace{2pt}
	    \caption{Architecture of the sub-networks in our framework.} 
	    \label{fig:subnetwork}
	
    \end{figure}

	\paragraph{\textbf{Segmentation Network.}}
	It is essential to build upon a good baseline model to achieve high-quality segmentation results. Our segmentation network follows the spirit of attention U-Net architecture \cite{oktay2018attention}. In encoder network, we keep the convolution layer as the initial setting. We perform three maxpool operations totally. Dilated convolution is adopted after third maxpool operation to capture large  receptive field to alleviate loss of structural information. Inspired by \cite{zhao2017pyramid}, pyramid pooling module is introduced to generate multi-scale features to alleviate the variance of heart size over each patient. In decoder network, we perform three deconvolution operations totally. For further accurate segmentation results, attention gate (as the black dot shown in Fig.~\ref{fig:subnetwork}(a)) is utilized to learn to focus on ventricles and myocardium structures. In attention gate, the features in the encoder part (as the blue rectangle shown in Fig.~\ref{fig:subnetwork}(a)) and decoder part (as the gray rectangle shown in Fig.~\ref{fig:subnetwork}(a)) are first squeezed with $3\times3$ convolution layer along the channel direction respectively and then added together. After that, we squeeze the features to single channel feature map to form structure attention with $1\times1$ convolution layer and generate final feature maps by dot product. Finally, we use $1\times1$ convolution layer with four output channels followed by the $sigmoid$ activation function to generate the probability maps. To save computational resources, we share the network with the same parameters between source and target domain.
	\vspace{-0.1cm}	
    \vspace{-0.9cm}		
	\paragraph{\textbf{Group-wise Feature Recalibration Module.}}
	Before we perform group-wise feature recalibration operation, different size features from segmentation network above are expanded and concatenated via upsampling and concatenating operations. The the features are send to GFRM. Our GFRM follows the spirit of \cite{roy2018concurrent}. Different from the above method, we divide features into four groups corresponding to the segmentation categories to focus on specific heart structures and we recalibrate features in each group (as shown in Fig.~\ref{fig:subnetwork}(b)). GFRM consists of two parts: channel attention part and spatial attention part. In channel attention part, we first squeeze global spatial information with global average pooling and fully connection layer. Then, we can generate the channel-wise attention features by simple dot product. In the spatial attention part, we first squeeze channel information with $1\times1$ convolution layer. Then, we can obtain the spatial-wise attention features by simple dot product. The features from channel attention part are added with the features from spatial attention part to generate group-wise recalibrated features. Finally, the features from each group are concatenated to generate final recalibrated features.
	\vspace{-0.1cm}	
	\paragraph{\textbf{Discriminator.}}
	The feature-level and mask-level discriminator are based on the multi-level features from GFRM and predicted mask results. We use PatchGAN as our discriminator \cite{isola2017image}. The network consists of $3$ convolution layers with stride of $2$ and $2$ convolution layers with stride of $1$. The kernel size of all convolution layers is $4\times4$ and the corresponding channel number is ${64, 128, 256, 256, 1}$. Except for the last layer, each convolution layer is followed by a leaky ReLU parameterized by $0.2$.
	\vspace{-0.1cm}	
	\subsection{Hybrid Loss Function for Source data}
	Since the labels for source domain are available, we train the segmentation network with a hybrid loss. The vanilla cross-entropy loss with our unbalanced training data leads to low accuracy. We add the Jaccard loss \cite{berman2018lovasz} into our loss function. The training objective for source data is
	\begin{equation}
	\mathcal L_{ce}^{s}=-\mathbb{E}_{x_{s}\sim S}(\sum_{i=1}^{N_{s}}\sum_{c=1}^{C}y_{s,i,c}\log G(x_{s,i};\Theta_{g}))
	\end{equation}
	\begin{equation}
	\mathcal L_{jac}^{s}=-\mathbb{E}_{x_{s}\sim S}(\sum_{i=1}^{N_{s}}\sum_{c=1}^{C}\frac{y_{s,i,c} G(x_{s,i};\Theta_{g})}{y_{s,i,c}+G(x_{s,i};\Theta_{g})-y_{s,i,c}G(x_{s,i};\Theta_{g})})
	\end{equation}
	$S$ represents source domain; For each source image $x_{s}$, there is one corresponding annotation $y_{s}$; $N_s$ is the number of all source images; $\mathbb{E}_{x_{s}\sim S}$ means that all $x_{s}$ are from $S$; $C$ is the number of all categories; $G$ is segmentation network; $\Theta_{g}$ is the parameters of $G$; $y_{s,i,c}$ and $G(x_{s,i};\Theta_{g})$ mean the annotation and prediction vectors, respectively. For cross entropy loss, the imbalance of training data leads to a local optimum with inappropriate direction of gradient decreasing, especially in the early stage. The Jaccard loss effectively helps to avoid the local optimum due to its better perceptual quality and scale invariance \cite{berman2018lovasz}.
	\subsection{Adversarial Learning for Target Data}
	In the target domain, due to the lack of annotations, we leverage the adversarial learning to train the segmentation network by minimizing the discrepancy across the source and target domain. Domain adaptation based on both feature and output space is proved to be effective for heart structure segmentation \cite{dou2018pnp}. In our framework, we employ two discriminators. The features input to feature domain discriminator are selected empirically in \cite{dou2018pnp}. To overcome this problem, we propose the GFRM to leverage the full feature spectrum and automatically select prominent features in the feature space. In the segmentation network, each feature scale generates one output feature map in the same dimension via convolution and upsampling operations. The feature maps are further processed by the GFRM to highlight the prominent features and suppress the irrelevant ones. The combined feature maps are then fed to the feature discriminator network for the adversarial learning, where the losses are defined as
	\begin{equation}
	\begin{split}
    \mathcal L_{adv_{D_f}}=&-\mathbb{E}_{x_{s}\sim S}\log D_f(R(G(x_{s};\Theta_{g});\Theta_{r});\Theta_{d_f})\\
                            &-\mathbb{E}_{x_{t}\sim T}(1-\log D_f(R(G(x_{t};\Theta_{g});\Theta_{r});\Theta_{d_f}))
    \end{split}
    \end{equation}

    \begin{equation}
    \mathcal L_{adv_{G_f}}=-\mathbb{E}_{x_{t}\sim T}\log D_f(R(G(x_{t};\Theta_{g});\Theta_{r});\Theta_{d_f})
    \end{equation}$T$ represents target domain; where $x_{t}$ is target data; $\mathbb{E}_{x_{t}\sim T}$ means that all $x_{t}$ are from $T$; $R$ is the GFRM; $\Theta_{r}$ is the parameters of $R$; $D_f$ is the feature discriminator; $\Theta_{d_f}$ is the parameters of $D_f$.\par	
	In the output space, the segmentation results of target domain should be similar to the ones of source domain. To achieve this, we employ the adversarial learning technique in the output space, where the losses are defined as
	\begin{equation}
	\begin{split}
    \mathcal L_{adv_{D_m}}=&-\mathbb{E}_{x_{s}\sim S}\log D_m(G(x_{s};\Theta_{g});\Theta_{d_m})\\
                            &-\mathbb{E}_{x_{t}\sim T}(1-\log D_m(G(x_{t};\Theta_{g});\Theta_{d_m}))
    \end{split}
    \end{equation}

    \begin{equation}
    \mathcal L_{adv_{G_m}}=-\mathbb{E}_{x_{t}\sim T}\log D_f(G(x_{t};\Theta_{g});\Theta_{d_m})
    \end{equation}where $D_m$ is the mask discriminator; $\Theta_{d_m}$ is the parameters of $D_m$.\par
    Combined with the aforementioned loss, the full objective function
    \begin{equation}
    \begin{split}
    \mathcal L_{FULL}=&\lambda_{ce}\mathcal L_{ce}+\lambda_{jac}\mathcal L_{jac}+\lambda_{D_f}\mathcal L_{adv_{D_f}}\\
    &+\lambda_{G_f}\mathcal L_{adv_{g_f}}+\lambda_{D_m}\mathcal L_{adv_{D_m}}+\lambda_{G_m}\mathcal L_{adv_{g_m}}
    \end{split}
    \end{equation}

    \section{Experiment}
\begin{figure}[h]
    	    \centering
    	    \hspace{1pt}
    	    \subfloat[]{\includegraphics[width=0.155\textwidth]{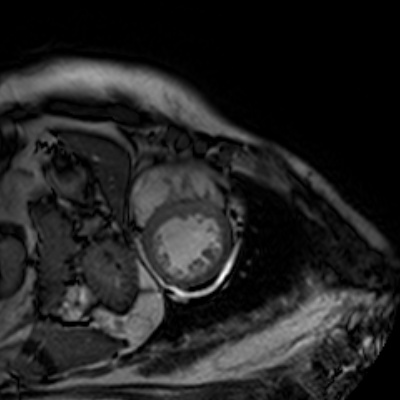}}
    	    \hspace{1pt}
    	    \subfloat[]{\includegraphics[width=0.155\textwidth]{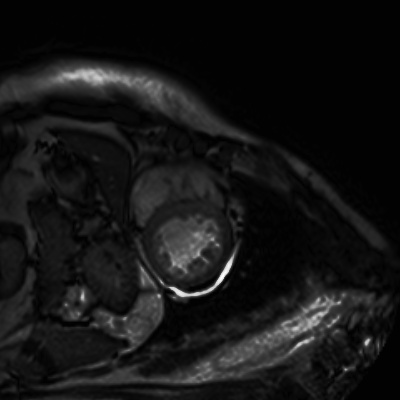}}		
    	    \hspace{1pt}
    	    \subfloat[]{\includegraphics[width=0.155\textwidth]{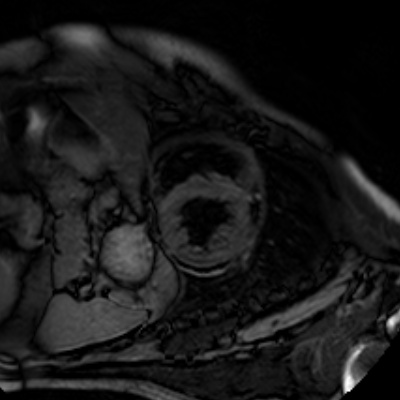}}	
    	    \hspace{1pt}
    	    \subfloat[]{\includegraphics[width=0.155\textwidth]{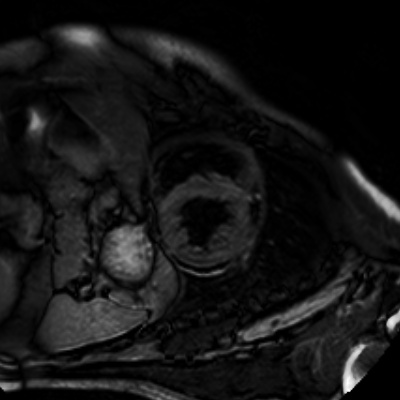}}
    	    \hspace{1pt}
    	    \subfloat[]{\includegraphics[width=0.155\textwidth]{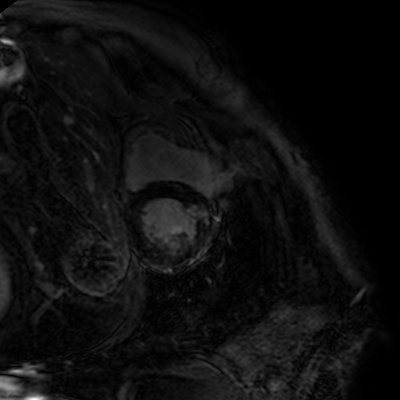}}
    	    \hspace{1pt}
    	    \subfloat[]{\includegraphics[width=0.155\textwidth]{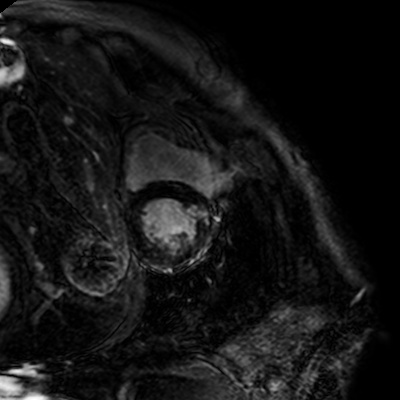}}    	
    	    \vspace{-5pt}
    	    \caption{\small Visual comparison for histogram match operation: (a) T2-weighted MRI. (b) T2-weighted MRI after histogram match. (c) bSSFP MRI. (d) bSSFP MRI after histogram match. (e) LGE MRI. (f) LGE MRI after histogram match.}
    	    \label{fig:pre-process}
        \end{figure}

    \paragraph{\textbf{Dataset.}}
    The validation of the proposed method is performed in the MS-CMRSeg Challenge 2019 dataset covering 45 patients. There are bSSFP, T2-weighted and LGE MRI sequences in each patient data. In one patient data, the slice number and annotation of three MRI modalities are different. We combine labeled bSSFP and T2-weighted MRI sequences as source data, and unlabeled LGE MRI sequences as target data. Experienced experts manually annotated the left
    ventricle (LV), right ventricle(RV) and myocardium (Myo) as ground truth. We pre-processing the data for domain adaptation. The data is resized and cropped to $400\times400$ in the center of each slice. In order to eliminate the inconsistency in appearance, we perform histogram match operation on both source and target data, as shown in Fig.~\ref{fig:pre-process}.

    \paragraph{\textbf{Implementation Details.}}
    In our experiments, we implement our whole network with PyTorch, using a standard PC with a single NVIDIA 1080Ti. To train the segmentation network, we use the Stochastic Gradient Descent (SGD) optimizer with Nesterov acceleration where the momentum is $0.9$ and the weight
         decay is $1e^-4$. The initial learning rate is set as $0.01$ and is decreased to $0.001$ after $80$ epochs. For training the both feature and mask discriminator, we use Adam optimizer with the fixed learning rate as $0.0002$. The weight decay is set as $5e^-5$. We totally trained $150$ epochs with a mini-batch size of $8$. We set $\lambda_{ce}$, $\lambda_{jac}$, $\lambda_{G_f}$, $\lambda_{D_f}$, $\lambda_{G_m}$ and $\lambda_{D_m}$ to $0.5$, $0.5$, $0.05$, $1.0$, $0.005$ and $1.0$. The training time cost only $5$ hours to converge.
    \begin{figure}[h]
    	\subfloat[ ]{\includegraphics[width=0.137\textwidth]{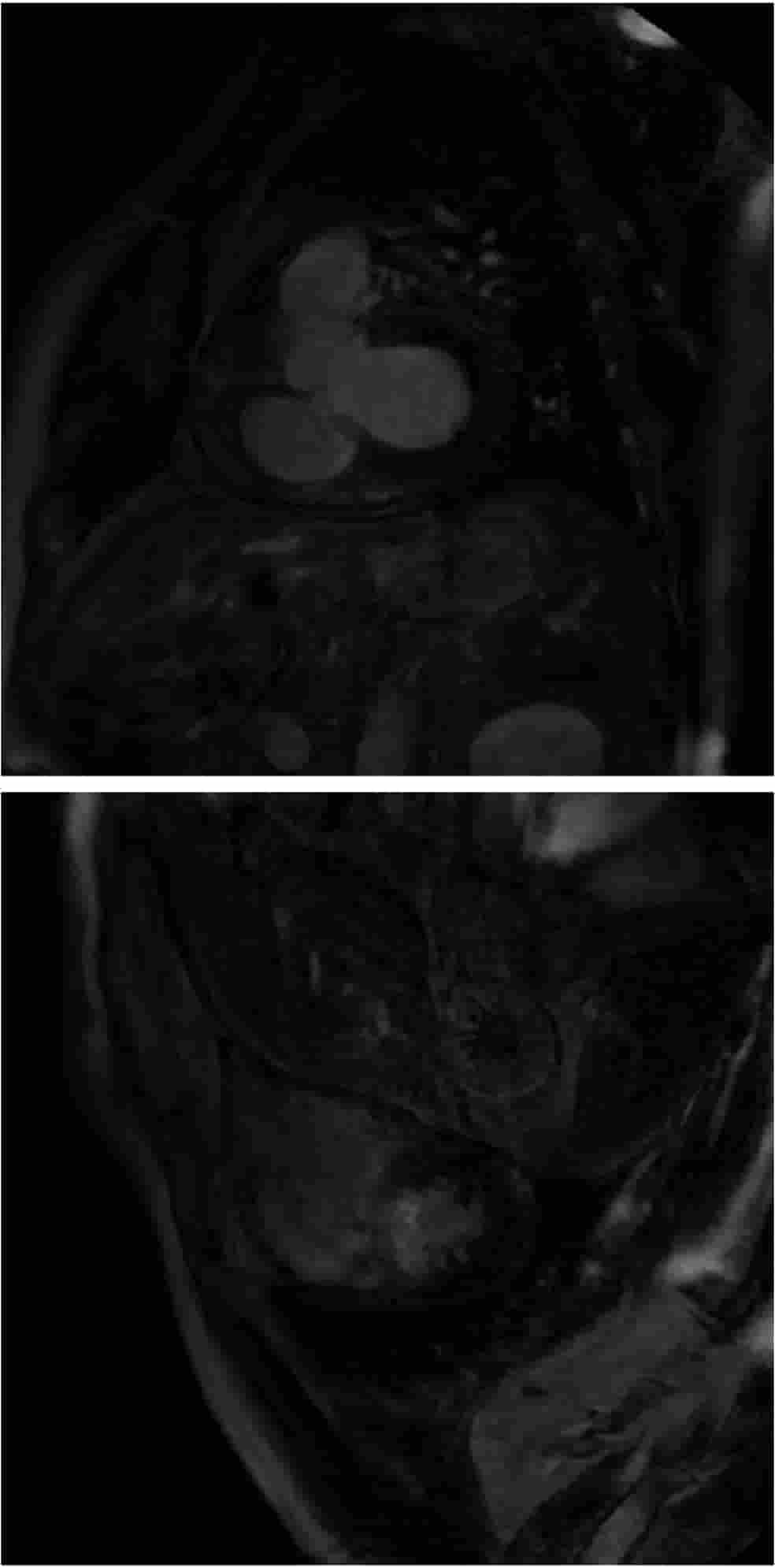}}
    	\hspace{0.1pt}
    	\subfloat[ ]{\includegraphics[width=0.137\textwidth]{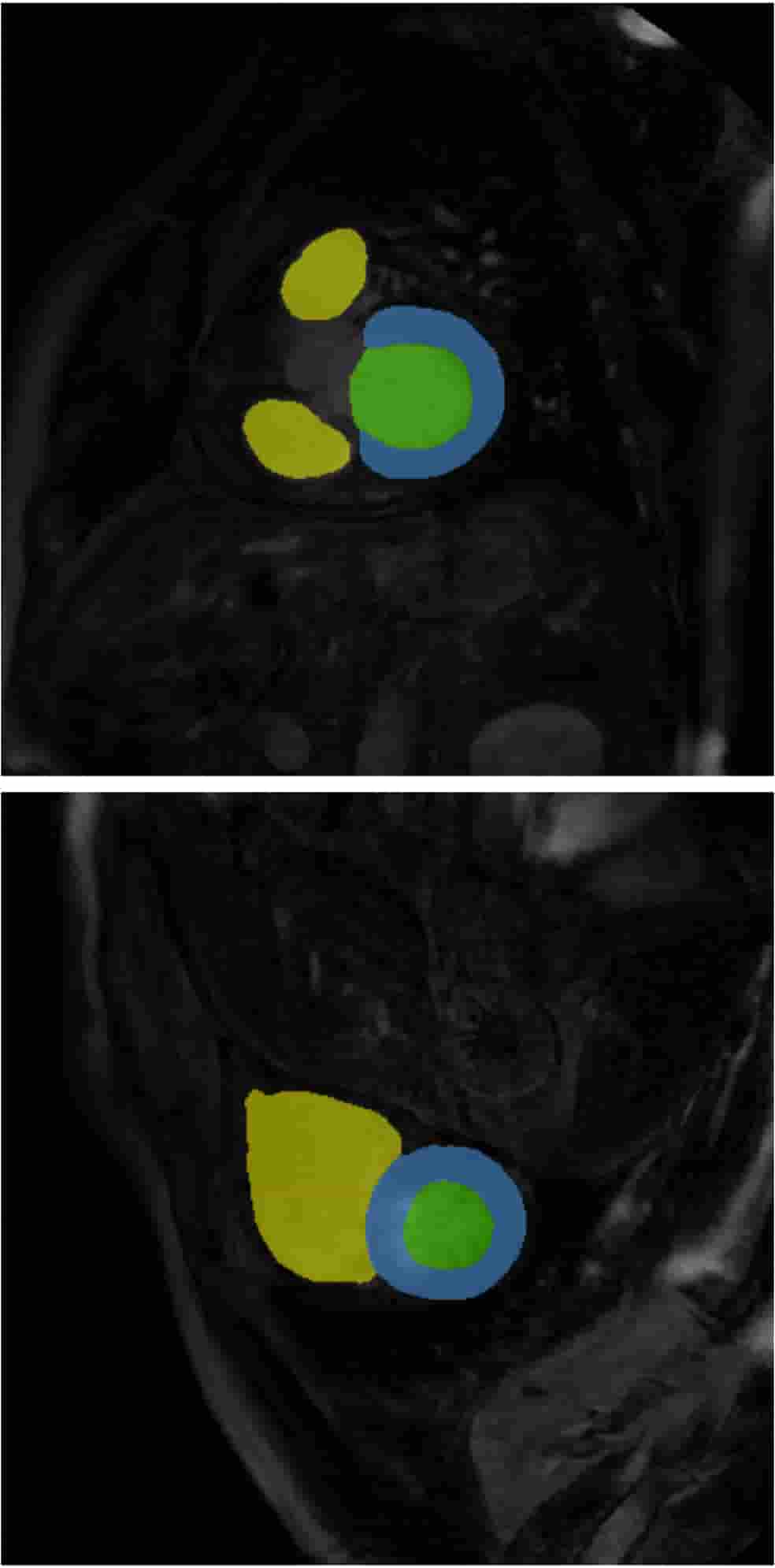}}
    	\hspace{0.1pt}
    	\subfloat[ ]{\includegraphics[width=0.137\textwidth]{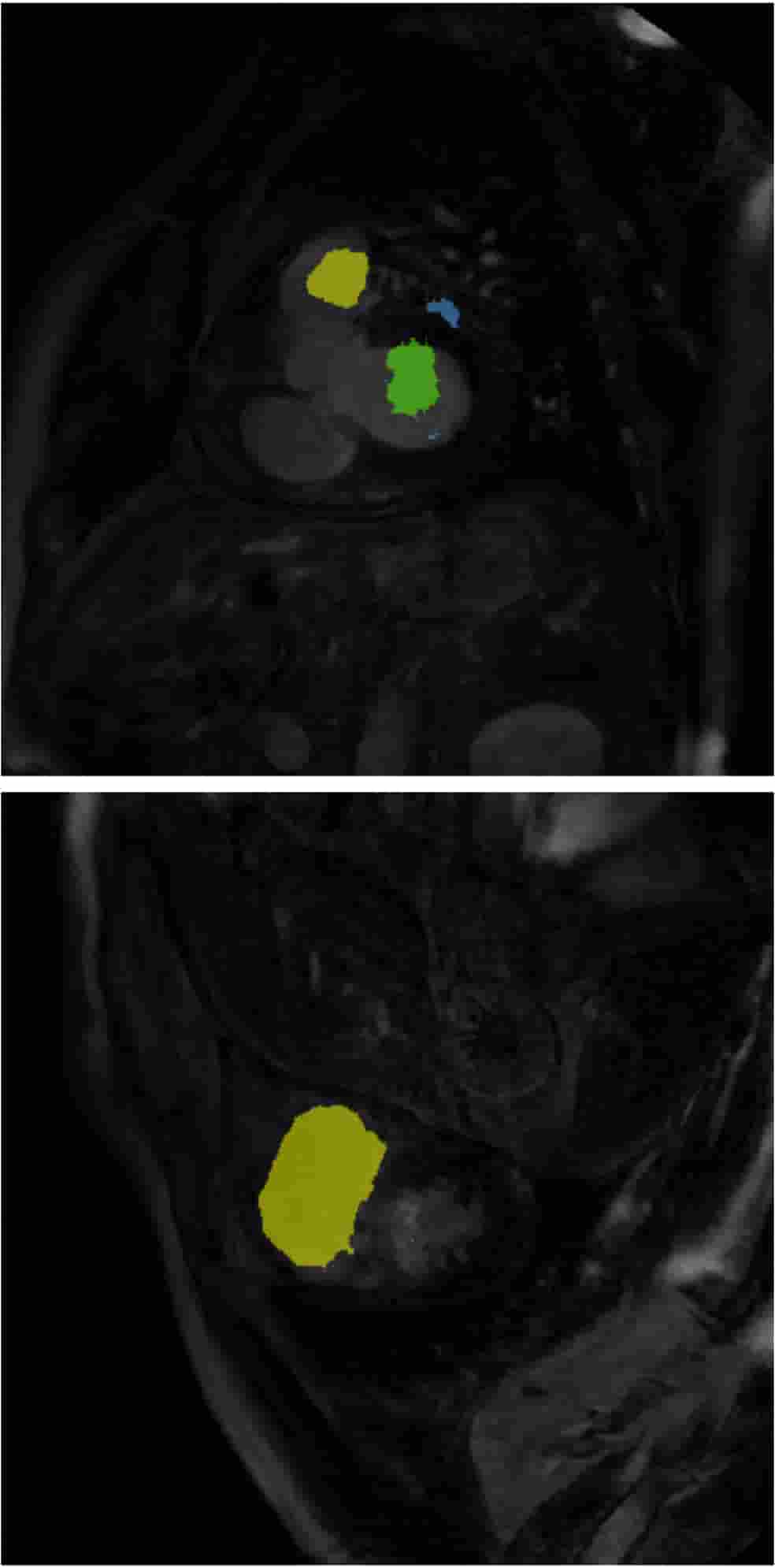}}
    	\hspace{0.1pt}
    	\subfloat[ ]{\includegraphics[width=0.137\textwidth]{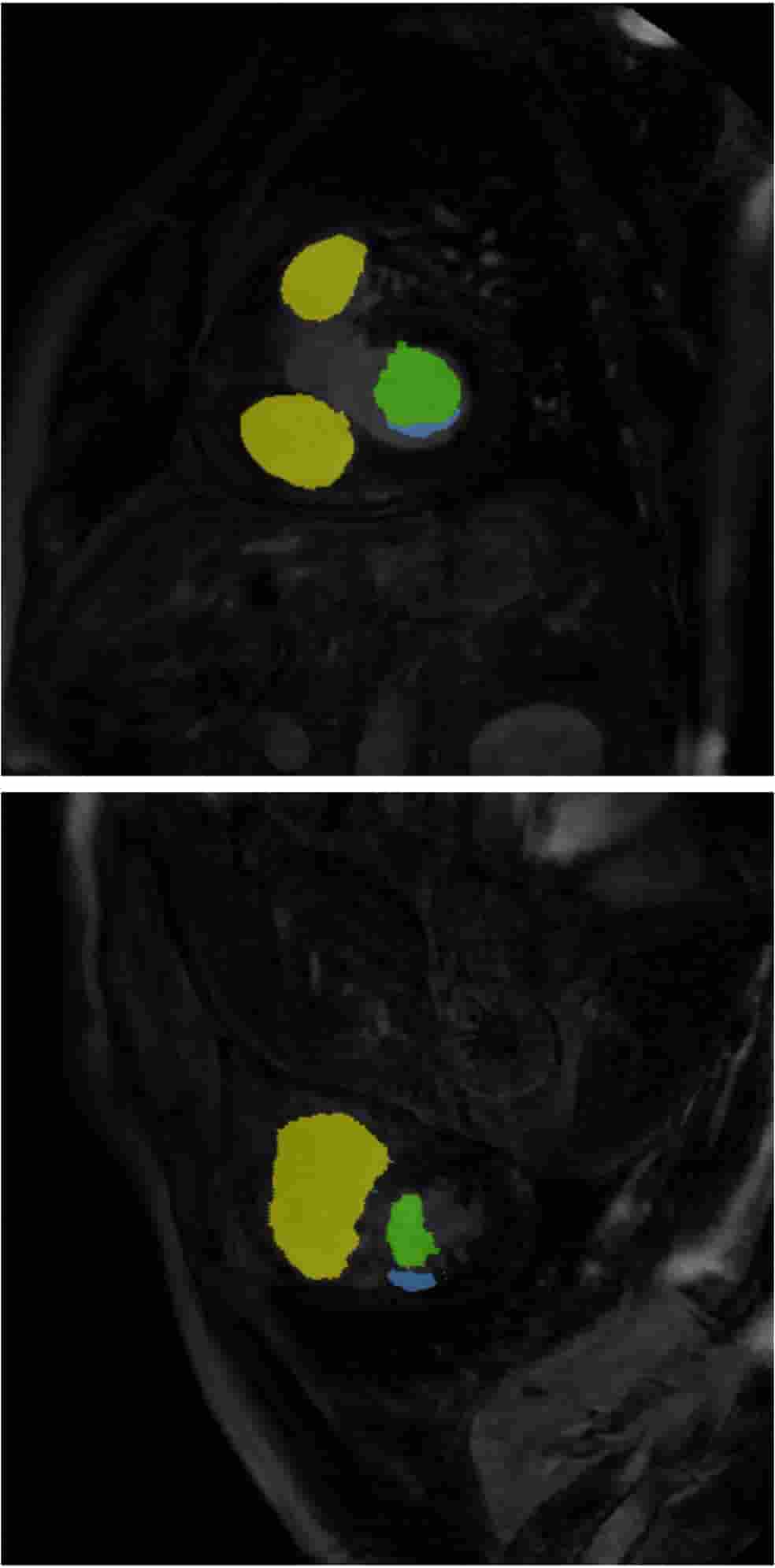}}
    	\hspace{0.1pt}
    	\subfloat[ ]{\includegraphics[width=0.137\textwidth]{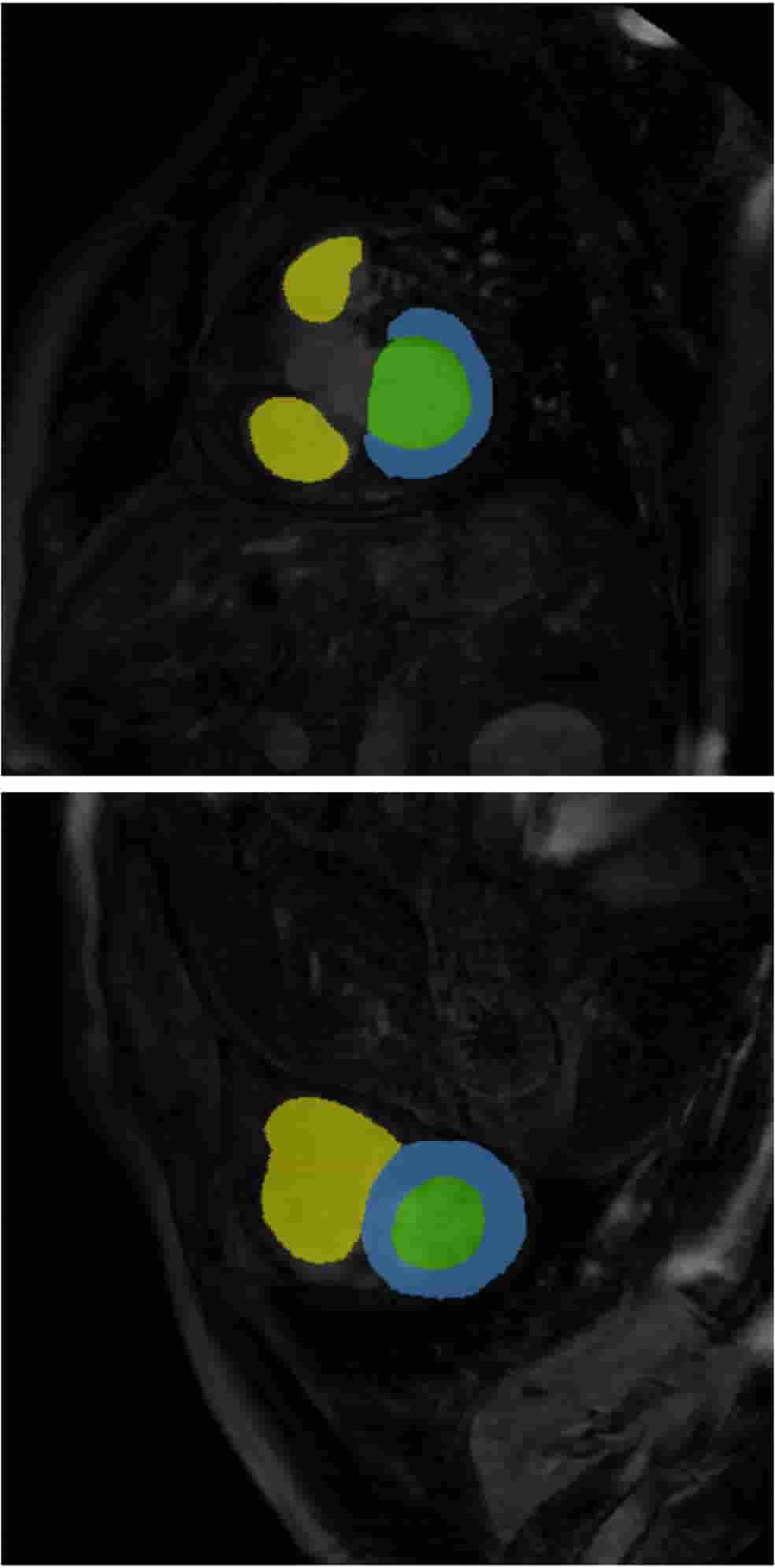}}
    	\hspace{0.1pt}
    	\subfloat[ ]{\includegraphics[width=0.137\textwidth]{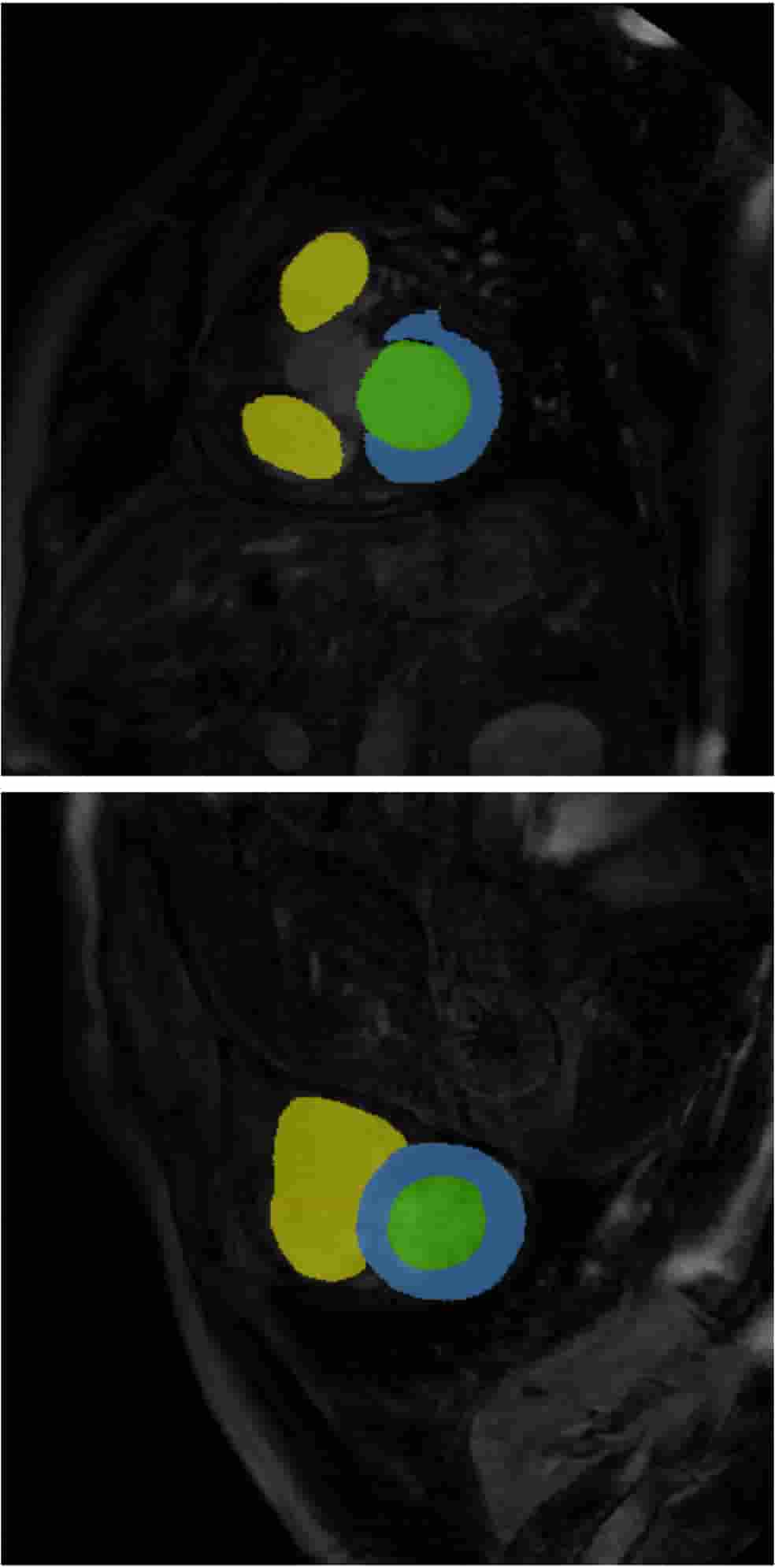}}
    	\hspace{0.1pt}
    	\subfloat[ ]{\includegraphics[width=0.137\textwidth]{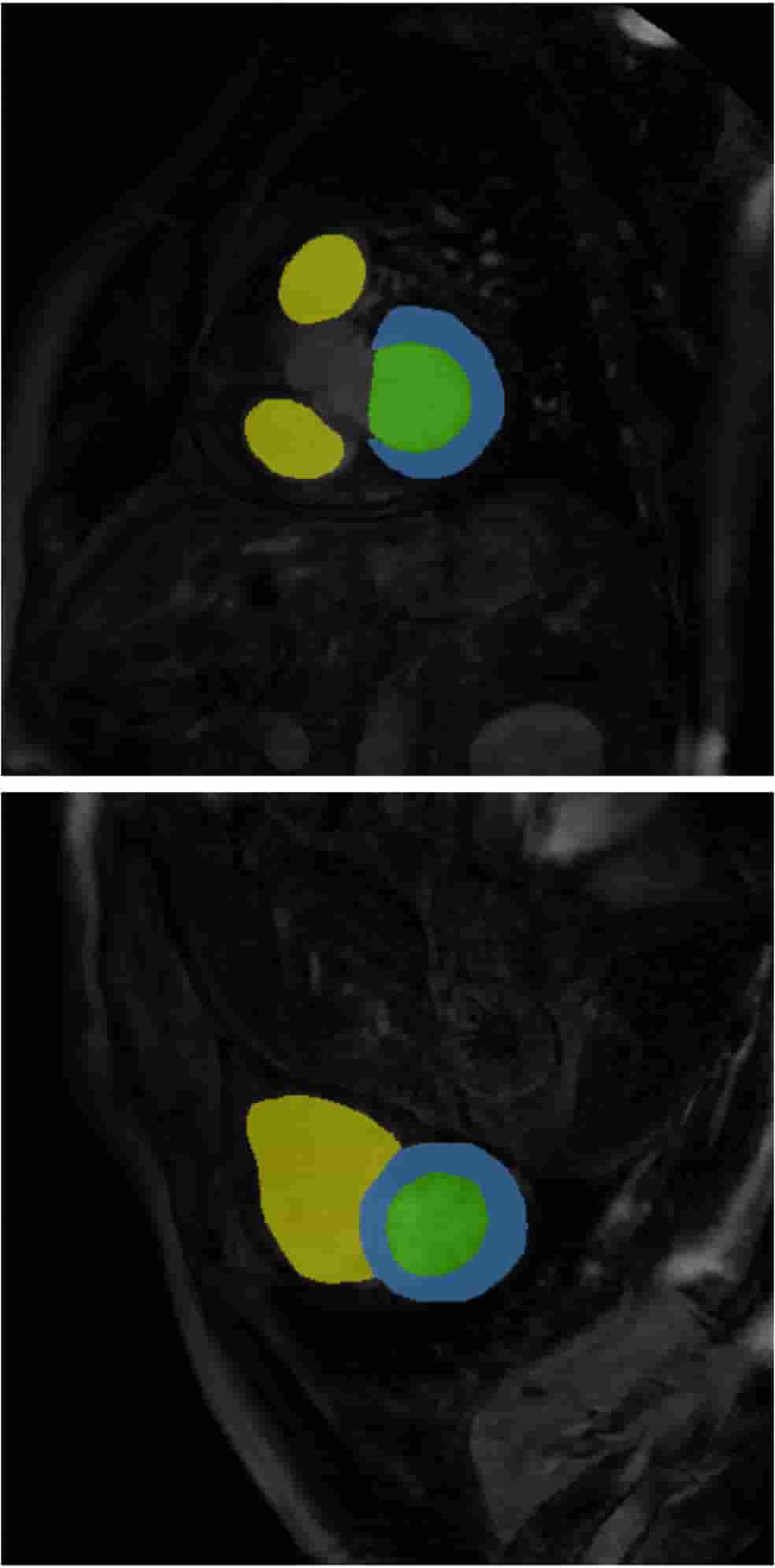}}
    	\hspace{0.1pt}
    	\vspace{-5pt}
    	\caption{\small Visual comparison for the LV, RV, and Myo segmentation results from ablation setting. (a) Original image from source domain. (b) Annotation. (c) S2T. (d) S2T+HM. (e) S2T+HM+MDA. (f) S2T+HM+MDA+FDA. (g) S2T+HM+MDA+FDA+GFRM.}
    	\label{fig:VisualComparison}
    \end{figure}
    \paragraph{\textbf{Quantitative and Qualitative Analysis.}}
    In order to verify the effectiveness of the proposed method, we adopt Dice coefficient (DSC), Jaccard coefficient (Jac) for further evaluation. We first trained segmentation network on the source data and then test on the target data (S2T). The results in Table 1 shows that the mean Dice in S2T is too slow. As we can see, our method can promote about $36.09\%$ in DSC and $38.38\%$ in Jac than S2T, which indicates that our method can alleviate dataset shift across different domains.\par
    In addition, we examine the effect of the histogram match operation (HM), mask-level adversarial learning (MDA), feature-level adversarial learning (FDA) and GFRM on the performance in the target domain. The result of the ablation study in Table 1 shows that our proposed modules can achieve a better performance than S2T. Fig.~\ref{fig:VisualComparison} demonstrates that each proposed module can contribute to alleviate the domain misalignment.
    \vspace{-0.4cm}
    \begin{table}[H] \caption {Quantitative evaluation of our proposed methods} \label{table:quanti_metric}
    	\scriptsize
    	\centering
    	\begin{spacing}{1.5}
    	    \begin{tabular}{l|C{0.9cm}|C{0.85cm}|C{0.9cm}|C{0.85cm}|C{0.9cm}|C{0.85cm}|C{0.9cm}|C{0.85cm}}			
    		    \toprule[2pt]
    		    \multirow{2}{*}{\bf{Method}} & \multicolumn{2}{C{2cm}|}{\bf{LV}} & \multicolumn{2}{C{1.5cm}|}{\bf{RV}} & \multicolumn{2}{C{1.5cm}|}{\bf{Myo}} & \multicolumn{2}{C{1.5cm}}{\bf{Mean}}\\
    		    \cline{2-9}		
    	        &DSC[\%] 	&Jac[\%]  	 	&DSC[\%] 	&Jac[\%]   	&DSC[\%] 	&Jac[\%]          &DSC[\%] 	&Jac[\%]\\
    		    \cline{0-1}
    		    \cline{2-9}
    		    \cline{0-1}
    		    \cline{2-9}
    		    S2T &50.01 &37.41  &66.72 &51.03 &31.69 &21.88&49.47 &36.78\\
    	       	S2T+HM &59.80 &47.66  &76.02 & 62.98 &38.13 &26.73&57.98 &45.79\\
    		    S2T+HM+MDA &85.67 &75.48  &86.19 &75.89 &75.35 &60.70&82.40 &70.69\\
    		    S2T+HM+MDA+FDA &88.43 &79.68  &85.70 &75.14 &78.43 &64.57&84.19 &73.13\\
    		    S2T+HM+MDA+FDA+GFRM &\textbf{89.33} &\textbf{81.15} & \textbf{87.17} &\textbf{77.29} &\textbf{80.17} &\textbf{67.04}&\textbf{85.56}&\textbf{75.16}\\
    		    \hline
    		    \toprule[2pt]
    	    \end{tabular}
        \end{spacing}
    \end{table}
    \vspace{-1.0cm}
	\section{Conclusion}
    In this paper, we proposed an unsupervised domain alignment method for left ventricle (LV), right ventricle(RV) and myocardium (Myo) segmentation from different cardiac MR sequences. We first introduced a segmentation network with hybrid segmentation loss to generate accurate prediction. We alleviate the dataset shift across different domains by leveraging the adversarial learning in both feature and output spaces. The proposed GFRM can enforce the fine-grained semantic-level feature alignment that matching features from different networks but with the same class label. Experiments show that the proposed method can achieve competitive results.
    \section{Acknowledgments}This work was supported in part by the National Natural Science Foundation of China under Grants 61571382, 81671766, 61571005, 81671674, 61671309 and U1605252, in part by the Fundamental Research Funds for the Central Universities under Grants 20720160075 and 20720180059, in part by the CCF-Tencent open fund, and the Natural Science Foundation of Fujian Province of China (No.2017J01126).

	%
	\bibliographystyle{splncs03}
	\bibliography{heart}
\end{document}